\documentclass[twocolumn,showpacs,preprintnumbers,aps]{revtex4}

\newcommand{\BE}{\begin{equation}}
\newcommand{\EE}{\end{equation}}
\newcommand{\BA}{\begin{eqnarray}}
\newcommand{\EA}{\end{eqnarray}}

\newcommand{\BW}{\begin{widetext}}
\newcommand{\EW}{\end{widetext}}

\input epsf.tex

\begin{document}


\title{Band gaps and localization of water waves over one-dimensional topographical bottoms}
\author{Zhong An$^1$}\author{Zhen
Ye$^2$}\affiliation{$^1$Department of Physics, Fudan University,
Shanghai, China\\ $^2$Department of Physics, National Central
University, Chungli, Taiwan}
\date{October 1, 2003}

\begin{abstract}

In this paper, the phenomenon of band gaps and Anderson
localization of water waves over one-dimensional periodic and
random bottoms is investigated by the transfer matrix method. The
results indicate that the range of localization in random bottoms
can be coincident with the band gaps for the corresponding
periodic bottoms. Inside the gap or localization regime, a
collective behavior of water waves appears. The results are also
compared with acoustic and optical situations.

\end{abstract}


\maketitle

When propagating through structured media, waves will be multiply
reflected or scattered, leading to many interesting phenomena such
as band gaps in periodic structures \cite{Mermin} and localization
in disordered media\cite{Anderson}. Within a band gap, waves are
evanescent; when localized, they remain confined in space until
dissipated. The phenomenon of band gaps and localization has been
both extensively and intensively studied for electronic,
electromagnetic, and acoustic systems. A great body of literature
is available, as summarized in Ref.~\cite{Sheng}.

The propagation of waters through underwater structures has also
been widely studied, because of its importance in a number of
coastal engineering problems. In particular, the consideration of
band gaps has been recently applied to water wave systems
\cite{Hare,Chou,Nature,APL}. Some of the advances have been
reviewed, for example, by McIver \cite{McIver}. On one hand, the
most recent experiment has used water waves to demonstrate the
phenomenon of Bloch waves as a result of the modulation by
periodic bottom structures \cite{Nature}. On the other hand, the
possible band gaps have been recently proposed for water waves
propagation through arrays of rigid cylinders that regularly stand
in the water \cite{APL}.

Relatively speaking, the more intriguing concept of Anderson
localization remains less touched in the context of water waves.
Although the earlier attempts do show that the localization
phenomenon is possible for water waves \cite{JFM}, a few important
issues have not been discussed. These issues include, for example,
the statistical properties of localization, the phase behavior of
the localized states, single parameter scaling and universality,
and the relation between localization and band gaps of
corresponding periodic situations. Recent research \cite{Deych,LY}
indicates that these issues are essential in discerning
localization effects. As such, it might be desirable to consider
the localization of water waves further.

In the present Letter, we wish to study the localization of water
waves in one-dimensional randomly structured bottoms, and its
relation with band gaps of the corresponding periodic bottoms. The
phase behavior of the water waves in the presence of localization
will also be discussed.

\begin{figure}[hbt]
\epsfxsize=2.5in\epsffile{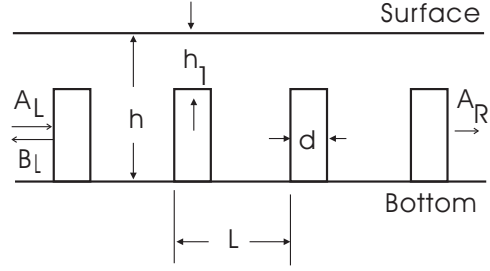} \smallskip \caption{Conceptual
layouts.} \label{fig1}
\end{figure}

The system we consider is illustrated in Fig.~\ref{fig1}. Here is
shown that there are $N$ identical steps of width $d$ periodically
placed on the bottom. The periodicity is $L$. The depth of water
is $h$, and the depth of the steps is $h_1$. We suppose that the
surface waves propagate from left to the right. The disorders are
introduced as follows. The degree of disorder is measured as the
percent of random displacement of the steps within a range around
their original regular positions with regard to the periodic
constant. Obviously the maximum (complete) randomness is $1 -
d/L$; half random (order) is thus $1/2 - d/(2L)$.

The wavenumbers in the water ($k$) and over the steps ($k_1$) are
determined as \cite{Ye} \BE \frac{\omega^2}{\omega_0^2} =
kL\tanh(kh), \ \mbox{and} \ \frac{\omega^2}{\omega_0^2} =
k_1L\tanh(k_1h_1),\EE where $\omega_0 = gL$. In the simulation,
all lengths are scaled by $L$, and the frequency is scaled by
$\omega_0$.

The waves on the left and right end of the step array can be
expressed in the matrix form \BE \left( \begin{array}{c}A_L e^{ikx}\\
B_Le^{-ikx}\end{array}\right), \ \mbox{and} \
\left( \begin{array}{c}A_R e^{ikx}\\
B_R e^{-ikx} \end{array}\right).\EE Clearly, $A_L$ is the incident
wave, $A_R$ the outgoing wave, and $B_L$ the reflected wave. $B_R$
is zero since there is no wave coming from the right. For a unit
plane wave incidence, $A_L =1$.

By the standard transfer matrix method \cite{LY}, the coefficients
$A_R, A_L, B_L$ can be related by a transfer matrix. The
transmission coefficient is defined as $T =
\left|\frac{A_R}{A_L}\right|^2.$ For the periodic case, the field
$\eta$ can be written in the Bloch form, $\eta(x) = e^{iKx}u(x)$,
where $u(x)$ is a periodic function with the periodicity of the
structure. Then the dispersion and band structure can be computed
from \BA \cos (KL) &=& \cos (k_1L (d/L)\cos(kL(L-d)/L) -\nonumber
\\ & & \cosh(2\xi)\sin(k_1L(d/L))\sin(kL(L-d)/L),\nonumber \\ \EA
where
$$\xi = \ln(q), \ \ \ \mbox{with} \ \ \ q^2 = \frac{k_1}{k} =
g_s^{-1} \ \ \ \Rightarrow \ \ \ \xi = -\frac{1}{2}\ln g_s.$$

\begin{figure}[hbt]
\epsfxsize=2.5in\epsffile{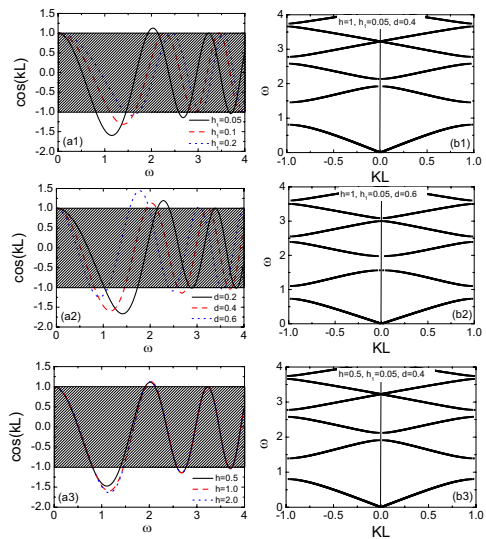} \smallskip \caption{The
dispersions and band structures versus frequency. Dispersions:
(a1) Variation of the step depth with $h = 1, d = 0.4$; (a2)
Variation of the step width $h = 1, h_1 = 0.05$; (a3) Variation of
the water depth with $h_1 = 0.05, d = 0.04$. The variation of the
band structures against $d$ and  $h$: ((b1) $h =1, h_1 = 0.5, d =
0.4$; (b2) $h = 1, h_1 =0.05, d=0.6$; (b3) $h=0.5, h_1 = 0.05,
d=0.4$.} \label{fig2}
\end{figure}

Fig.~\ref{fig2} shows the dispersion and band structures for
periodic systems. Figs.~\ref{fig2}(a1), (a2) and (a3) present the
dependence of the dispersion on the variations of the depth and
the width of the steps, and the water depth. The curves within or
outside the dark areas refer to the passing or forbidden bands
respectively. We observe the following from Fig.~\ref{fig2}. (1)
There are band gaps for water systems. But with increasing $h_1$,
the band gaps become narrower, referring to (a1). (2) The band
gaps move towards lower frequencies by increasing the width of the
steps, as shown by (a2). (3) While the band structures are
sensitive to the physical structures of the steps, they are rather
insensitive to the variation of the water depth, as evidenced by
(a3). (4) The band gaps start to disappear as the frequency
increases. This is understandable. For high frequencies,
especially when $kh, kh_1 >> 1$, the dispersion relation becomes
$\omega^2 = gk$, therefore the structure of the bottom has less
and less effects. Some band structures are exemplified by (b1),
(b2) and (b3). In our simulation, we found that the step depth
$h_1$ is vital in determining the band structures, and
subsequently the localization.

\begin{figure}[hbt]
\epsfxsize=2.5in\epsffile{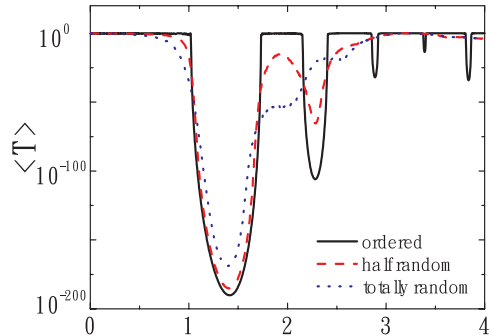} \smallskip
\caption{Transmission versus frequency for the ordered, half
ordered, and completely random cases: $h = 1, h_1 = 0.05, d =
0.2$. Here $\langle\cdot\rangle$ refers to the average over the
random configurations, and totally 100 steps are consideration in
the computation.} \label{fig3}
\end{figure}

\begin{figure}[hbt]
\epsfxsize=2.5in\epsffile{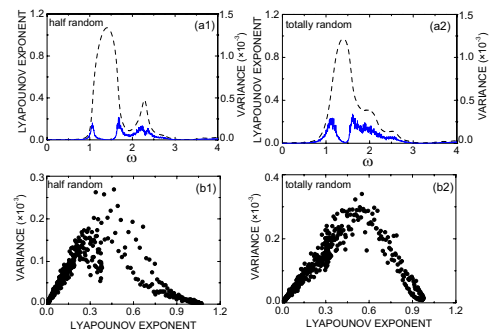} \smallskip \caption{The
behavior of the Lyapunov exponent and its variance. Here $h=1, h_1
= 0.05, d = 0.2.$} \label{fig4}
\end{figure}

In Fig.~\ref{fig3}, we show the transmission versus frequency for
different randomness. As a comparison, the transmission for the
ordered case is also shown. The results indicate the following.
(1) There are well defined inhibited transmission ranges for the
ordered case, coincident with the band gaps shown in
Fig.~\ref{fig2}. And these ranges decrease with frequency, until
disappear. The degree of inhibition more or less decreases with
frequency. These features differ from those in optical and
acoustic systems \cite{LY,Maradudin}. (2) The disorder tends to
reduce the transmission for the mid range of frequency, but it has
less influences for low or high frequencies, implying that the
scattering by the steps is weak in these ranges. The inhibition in
the presence of disorders is an indication and measure of
localization. (3) Within the band gaps, the disorder tends to
enhance the wave propagation, a feature that has been also
discovered previously in the optical and acoustic systems
\cite{LY,Maradudin}. But different from these systems, the
inhibition starts to decease as frequency increases. We also
notice that in some cases the strong localization is coincident
with the band gaps, particularly for the first two band gaps. (4)
The level of localization does not necessarily depend on the
degree of disorders. For example, within the first band gap, the
wave is more localized for the half random than the completely
random case.

\begin{figure}[hbt]
\epsfxsize=2.5in\epsffile{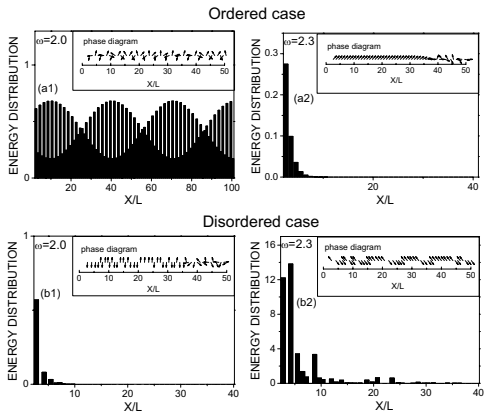} \smallskip \caption{Energy
distribution and the phase behavior (insets) for the ordered and
totally random cases. The structure parameters are the same as in
Fig.~\ref{fig4}.} \label{fig5}
\end{figure}

We have also studied the statistical properties of localization by
considering the Lyapounov exponent (LE) $\gamma =
\lim_{N\rightarrow \infty} \langle -\frac{1}{N}\ln(T) \rangle$,
where $N$ is the number of steps, and its variance as a function
of frequency. LE characterizes the degree of localization, and its
variance signifies the transition behavior. The results are shown
in Fig.~\ref{fig4}. The parameters are indicated in the caption.
Here we see that behavior of LE mimics the hand structures,
particularly for weak disorders. Similar to the optical case
\cite{Deych}, but contrary to the acoustic case \cite{LY}, there
are two double maxima for the variance inside the gaps. This
feature is more prominent in the low frequency bands. Different
from both optical and acoustic cases, the double peak feature in
the variance does not disappear even in the complete random case,
depicted by (a2). We also plot LE versus its variance in
Fig.~\ref{fig4}. Here even in the totally random situation, we do
not observe the linear dependence between LE and its variance in
contrast to the expectation from the scaling analysis of
localization \cite{SST}.

The energy distribution and the phase behavior of water waves in
the ordered and disordered cases are shown in Fig.~\ref{fig5}. The
energy is defined as the modulus of the waves, and the phase
$\theta$ is defined from $\eta = |\eta|e^{i\theta}$. To show the
phase behavior, we associate a phase vector to the phase as
$\vec{v} = (\cos\theta, \sin\theta)$ and the phase vector is
plotted in two-dimensions. Two frequencies are chosen to be inside
and outside the second gap, referring to Fig.~\ref{fig3}.
Comparing the ordered and disordered cases, we observe clearly
that the energy or the wave is indeed confined by the disorder to
the site of transmission. But the decay of the energy along the
path does not follow the exponential feature. Here we see, within
the gaps in the ordered case or when the localization occurs, the
phase vectors at different space points tends to point to the same
or the opposite directions. Such a collective behavior of water
waves has not been discussed before. At the end of the step
arrays, there is some disorientation in the pointing directions of
the phase vectors. This is due to the finite size effect.

In summary, we have applied the concept of band gaps and
localization to water surface waves in a one-dimensional system.
The statistical properties of localization and their relation with
corresponding band gaps are studied. It is shown that localization
is related to a coherent behavior of the system.

{\bf Acknowledgements} The work received support from Natural
Science Foundation of China (No.10204005), and the Shanghai Bai Yu
Lan fund (ZY).


\end{document}